\documentclass[envcountsame,orivec]{llncs}
\usepackage{amssymb}
\usepackage{stmaryrd}
\usepackage{xspace}
\usepackage{cmtt}
\usepackage{at}
\usepackage[ligature,reserved]{semantic}
\usepackage{ifthen}
\usepackage{calc}
\usepackage{mathenv}
\usepackage{mathpartir}
\usepackage{prooftree}
\usepackage{verbdef}
\usepackage{graphicx}
\usepackage{prettyref}
\usepackage{smartref}
\usepackage{theorem}
\usepackage{QED}
\usepackage{wasysym}
\usepackage{float}
\usepackage{placeins}
\usepackage{listings}

\DeclareMathAlphabet{\mathpzc}{OT1}{pzc}{m}{it}
\DeclareMathAlphabet{\mathsl}{OT1}{cmr}{m}{sl}

\newcommand{\magicwand}{\mathrel{\mbox{$\hspace*{-0.03em}\mathord{-}\hspace*{-0.66em}\mathord{-}\hspace*{-0.36em}\mathord{*}$\hspace*{-0.005em}}}}

\newcommand\Perp{\protect\mathpalette{\protect\PerP}{\perp}}
\def\PerP#1#2{\mathrel{\rlap{$#1#2$}\mkern2mu{#1#2}}}

\newcommand{\loPrecLogConSpc}[1]{\mathrel{#1}}
\newcommand{\hiPrecLogConSpc}[1]{\mathbin{#1}}
\newcommand{\expRelSpc}[1]{\mathord{#1}} 
\newcommand{\expOpSpc}[1]{\mathord{#1}} 
\newcommand{\semRelSpc}[1]{\mathbin{#1}} 
\newcommand{\semOpSpc}[1]{\mathord{#1}} 
\mathlig{//->}{\expRelSpc{\not\hookrightarrow}} 
\mathlig{--->}{\loPrecLogConSpc{\longrightarrow}} 
\mathlig{<---}{\loPrecLogConSpc{\longleftarrow}} 
\mathlig{-->}{\loPrecLogConSpc{\rightarrow}} 
\mathlig{-/>}{\loPrecLogConSpc{\nrightarrow}} 
\mathlig{<--}{\loPrecLogConSpc{\leftarrow}} 
\mathlig{</-}{\loPrecLogConSpc{\nleftarrow}} 
\mathlig{<->}{\loPrecLogConSpc{\leftrightarrow}} 
\mathlig{</>}{\loPrecLogConSpc{\nleftrightarrow}} 
\mathlig{--*}{\loPrecLogConSpc{\magicwand}} 
\mathlig{~/>}{\not\leadsto} 
\mathlig{-.>}{\dashrightarrow}
\mathlig{>->}{\rightarrowtail}
\mathlig{|->}{\expRelSpc{\mapsto}} 
\mathlig{/->}{\expRelSpc{\hookrightarrow}} 
\mathlig{|=>}{\Mapsto} 
\mathlig{==>}{\loPrecLogConSpc{\Rightarrow}} 
\mathlig{<==}{\expRelSpc{\Leftarrow}} 
\mathlig{<=>}{\expRelSpc{\Leftrightarrow}} 
\mathlig{--`}{\rightharpoonup} 
\mathlig{.|.}{\semRelSpc{\perp}} 
\mathlig{./.}{\semRelSpc{\not\perp}} 
\mathlig{|||}{\semOpSpc{\parallel}} 
\mathlig{===}{\mathrel{\equiv}} 
\mathlig{=/=}{\mathrel{\not\equiv}} 
\mathlig{|/-}{\nvdash} 
\mathlig{|/=}{\nvDash} 
\mathlig{...}{\cdots}
\mathlig{,.,}{,\ldots,}
\mathlig{..,}{\ldots,}
\mathlig{,..}{,\ldots}
\mathlig{::=}{\mathrel{::=}} 
\mathlig{:==}{\mathpunct{:\equiv}} 
\mathlig{!!!}{\mathord{!}\mathord{!}} 
\mathlig{->}{\expOpSpc{\shortrightarrow}} 
\mathlig{~>}{\leadsto} 
\mathlig{/|}{\hiPrecLogConSpc{\wedge}} 
\mathlig{|/}{\hiPrecLogConSpc{\vee}} 
\mathlig{::}{\loPrecLogConSpc{:}} 
\mathlig{??}{\loPrecLogConSpc{?}} 
\mathlig{++}{\expOpSpc{+}} 
\mathlig{--}{\expOpSpc{-}} 
\mathlig{==}{\expRelSpc{=}} 
\mathlig{!=}{\expRelSpc\neq} 
\mathlig{>>}{\expRelSpc{>}} 
\mathlig{<=}{\expRelSpc{\leq}} 
\mathlig{=>}{\expRelSpc{\Rightarrow}} 
\mathlig{||}{\mathrel{\parallel}} 
\mathlig{**}{\semOpSpc{*}} 
\mathlig{!!}{\mathord{!}} 
\mathlig{|-}{\vdash} 
\mathlig{-|}{\dashv} 
\mathlig{|=}{\vDash} 
\mathlig{|>}{\mathrel{\rhd}}
\mathlig{|[}{\llbracket} 
\mathlig{]|}{\rrbracket} 
\mathlig{|,}{\lfloor}
\mathlig{,|}{\rfloor}
\mathlig{|`}{\lceil}
\mathlig{`|}{\rceil}
\mathlig{~=}{\mathrel{\cong}} 
\mathlig{,.}{\mathop{,}} 
\mathlig{;.}{\mathop{;}} 
\mathlig{,,}{\mathbin{,}} 
\mathlig{;;}{\mathbin{;}} 
\mathlig{|.}{\,\mathord{\mid}\,} 
\mathlig{:=}{\mathpunct{:=}} 
\mathlig{;}{\loPrecLogConSpc{\brokenvert}} 
\mathlig{*}{\hiPrecLogConSpc{*}} 
\mathlig{!}{\neg} 
\mathlig{~}{\mathord{\sim}} 
\mathlig{|}{\mid} 
\mathlig{:}{\colon} 

\newcommand{\gen}[4][]{#2(#3).\ifthenelse{\equal{#1}{}}{\,}{\E[#1]}#4}

\newcommand{\A}[1][]{\forall \ifthenelse{\equal{#1}{}}{}{\mbox{$#1.$}\,}}
\newcommand{\E}[1][]{\exists \ifthenelse{\equal{#1}{}}{}{\mbox{$#1.$}\,}}
\newcommand{\Am}[1][]{\forall_{\!*} \ifthenelse{\equal{#1}{}}{}{#1.\,}}
\newcommand{\Em}[1][]{\exists_{*} \ifthenelse{\equal{#1}{}}{}{#1.\,}}
\newcommand{\nA}[1][]{\not\forall \ifthenelse{\equal{#1}{}}{}{#1.\,}}
\newcommand{\nE}[1][]{\nexists \ifthenelse{\equal{#1}{}}{}{#1.\,}}

\newcommand{\al}[2][]{\lambda \mbox{$#2$}\ifthenelse{\equal{#1}{}}{}{^{#1}}.\,}

\newcommand{\ml}[2][]{\lambda_{*} \mbox{$#2$}\ifthenelse{\equal{#1}{}}{}{^{#1}}.\,}

\makeatletter
\@ifundefined{case}{
  
}{}
\makeatother

\newcommand{\spec}[3]{\{#1\}\;#2\;\{#3\}} 

\newcommand{\mlA}[1][]{
  \metalanglogicstyle{for\ all\;} \ifthenelse{\equal{#1}{}}{}{#1.\,}}
\newcommand{\mlE}[1][]{
  \metalanglogicstyle{there\ exists\;} \ifthenelse{\equal{#1}{}}{}{#1.\,}}
\newcommand{\lam}[2][]{\lambda \mbox{$#2$}\ifthenelse{\equal{#1}{}}{}{^{#1}}.\,}

\newcommand{\dlam}[2][]{\Pi \mbox{$#2$}\ifthenelse{\equal{#1}{}}{}{^{#1}}.\,} 

\renewcommand{\setminus}{\semOpSpc{\smallsetminus}} 
\newatcommand | [1]{\mathord{\mid}_{#1}} 

\newatcommand \ {\setminus}

\renewcommand{\u}{\cup}
\newcommand{\n}{\cap}

\newcommand{\disjoint}[2]{#1\pitchfork#2}
\newcommand{\set}[1]{\left\{#1\right\}}

\newlength{\strtlen}
\newlength{\strtdep}

\newsavebox{\Overlinebox} \newlength{\Overlinelen}

\newcommand{\fext}[2][]{[\ifthenelse{\equal{#1}{}}{}{#1|.}#2]}
\renewcommand{\to}{\mathord{\shortrightarrow}}

\newcommand{\stacklabel}[1]{\stackrel{\smash{\scriptscriptstyle \mathrm{#1}}}}
\newcommand{\Def}{\stacklabel{def}}

\newcommand{\asgreekstyle}{\mathrm} 
\newatcommand a {\ensuremath{\alpha}\xspace}
\newatcommand b {\ensuremath{\beta}\xspace}
\newatcommand c {\ensuremath{\xi}\xspace}
\newatcommand C {\ensuremath{\Xi}\xspace}
\newatcommand d {\ensuremath{\delta}\xspace}
\newatcommand D {\ensuremath{\Delta}\xspace}
\newatcommand e {\ensuremath{\varepsilon}\xspace}
\newatcommand E {\ensuremath{\asgreekstyle{E}}\xspace}
\newatcommand f {\ensuremath{\phi}\xspace}
\newatcommand F {\ensuremath{\Phi}\xspace}
\newatcommand g {\ensuremath{\gamma}\xspace}
\newatcommand G {\ensuremath{\Gamma}\xspace}
\newatcommand h {\ensuremath{\varrho}\xspace}
\newatcommand j {\ensuremath{\varsigma}\xspace}
\newatcommand k {\ensuremath{\kappa}\xspace}
\newatcommand l {\ensuremath{\ell}\xspace}
\newatcommand L {\ensuremath{\Lambda}\xspace}
\newatcommand m {\ensuremath{\mu}\xspace}
\newatcommand n {\ensuremath{\eta}\xspace}
\newatcommand o {\ensuremath{\theta}\xspace}
\newatcommand O {\ensuremath{\Theta}\xspace}
\newatcommand p {\ensuremath{\varphi}\xspace}
\newatcommand P {\ensuremath{\Pi}\xspace}
\newatcommand q {\ensuremath{\epsilon}\xspace}
\newatcommand Q {\ensuremath{\Phi}\xspace} 
\newatcommand r {\ensuremath{\rho}\xspace}
\newatcommand s {\ensuremath{\sigma}\xspace}
\newatcommand S {\ensuremath{\Sigma}\xspace}
\newatcommand t {\ensuremath{\tau}\xspace}
\newatcommand u {\ensuremath{\upsilon}\xspace}
\newatcommand U {\ensuremath{\Upsilon}\xspace}
\newatcommand v {\ensuremath{\nu}\xspace}
\newatcommand w {\ensuremath{\omega}\xspace}
\newatcommand W {\ensuremath{\Omega}\xspace}
\newatcommand x {\ensuremath{\chi}\xspace}
\newatcommand y {\ensuremath{\psi}\xspace}
\newatcommand Y {\ensuremath{\Psi}\xspace}
\newatcommand z {\ensuremath{\zeta}\xspace}

\newcommand{\smtt}[1]{\mbox{\small$\tt#1$}}

\newcommand{\metalangfuncstyle}[1]{\ensuremath{\mathsl{#1}}}
\reservestyle{\metalangfunc}{\metalangfuncstyle}
\metalangfunc{dom, fv, i, rng}

\newcommand{\metalangmvarstyle}[1]{\ensuremath{\mathit{#1}}}
\reservestyle{\metalangmvar}{\metalangmvarstyle}

\newcommand{\metalanglogicstyle}[1]{\ensuremath{\mathrm{#1}}}
\reservestyle{\metalanglogic}{\metalanglogicstyle}
\metalanglogic{always, and[\;and\;], def[\;defined], either[\;either\;],
  for[\;for\;], forany[\;for\; any\;], fresh[\;fresh], if[\;if\;],
  iff[\;iff\;], implies[\;implies\;], never, not[not\;], onlyif[\;only\ 
  if\;], or[\;or\;], otherwise, undef[\;undefined], when[when\;]}
\newlength{\thickspc} 
\newatcommand ! {\nolinebreak\hspace{-\thickspc}}

\newcommand{\semanticdomainstyle}[1]{%
  \ensuremath{\mathchoice%
    {\mbox{\normalfont#1}}%
    {\mbox{\normalfont#1}}%
    {\mbox{\normalfont\scriptsize#1}}%
    {\mbox{\normalfont\tiny#1}}}}
\reservestyle{\semanticdomain}{\semanticdomainstyle}
\semanticdomain{Heaps, Loc, Lval[L-val], Rval[R-val], Stacks, States, Val, Var, Wld, Set, Nat[\ensuremath{\mathbb{N}}]}

\newcommand{\semanticvaluestyle}[1]{\ensuremath{\mathrm{#1}}}
\reservestyle{\semanticvalue}{\semanticvaluestyle}
\semanticvalue{semp[\varnothing], ee[\<mu>], snil[nil], tt[\<au>], ff}

\newcommand{\proglangkeywordstyle}[1]{\mbox{\rm$\mtt{#1}$}}
\reservestyle{\proglangkeyword}{\proglangkeywordstyle}
\proglangkeyword{let[let\;], in[\;in\;], choose[choose\;],
  case[case\;], of[\;of\;], skip, P, V, goto, swap, new, free, disp[dispose]}

\newcommand{\proglangidentstyle}[1]{\ensuremath{\mathrm{#1}}}
\reservestyle{\proglangident}{\proglangidentstyle}

\newcommand{\syntacticpredicatestyle}[1]{\ensuremath{\mathsf{#1}}}
\reservestyle{\syntacticpredicate}{\syntacticpredicatestyle}
\syntacticpredicate{emp, false, ls, nil, true, int, az[\bot], mz[\Perp],
  au[true], mu[emp], aeb[\<au>], meb[\<mu>], ef[\emptyset]}

\reservestyle{\infrule}{\infrulelabel}
\infrule{}

\newcommand{\acronymstyle}[1]{\textsc{#1}}
\reservestyle{\acronym}{\acronymstyle}
\acronym{cps, ds, bi, sci}

\newcommand{\plks}{\proglangkeywordstyle}
\newcommand{\sps}{\syntacticpredicatestyle}
\newcommand{\itee}[3]{\sps{if}\;#1\;\sps{then}\;#2\;\sps{else}\;#3}
\newcommand{\ite}[3]{\plks{if}(#1)\;\{#2\}\;\plks{else}\;\{#3\}}
\newcommand{\while}[3][]{
 \plks{while}(#2)\ifthenelse{\equal{#1}{}}{}{\;[#1]\;}\{#3\}}

\newcommand{\with}[3]{\plks{with}\;#1\;\plks{when}\;#2\;\plks{do}\;#3}

\makeatletter\@ifundefined{text}{
  \newcommand{\text}[1]{\mbox{\normalfont#1}}
}\makeatother

\newsavebox{\tmpbox}
\newlength{\parindentsave}

\newlength{\hsmashlen}

\let\origae=\ae
\renewcommand{\ae}{\origae\xspace}

  {[\textsc{Aside:}}
  {]}

   {\begin{list}{}{\setlength{\labelwidth}{\leftmargin+\itemindent-\labelsep}%
         \setlength{\parsep}{0.25ex+\parskip}%
     }}%
   {\end{list}}

\renewcommand{\TirName}[1]{$#1$}
\renewcommand{\RefTirName}[1]{$#1$}
\newcommand{\infrulelabel}[1]{\mbox{\normalfont\small\textsc{#1}}}
\newcommand{\drule}[4][]
  {\inferrule*[lab=\<#2>,right=\ensuremath{#1}]{#3}{\textstyle#4}}

\renewcommand{\qedsymbol}{\scriptstyle{\boxempty}}

\makeatletter
  {\protect\@Proof%
   \protect\sloppy%
   \ifthenelse{\equal{#1} {}}%
     {}%
     {\hskip-0.75em\relax{\bf[#1]}\enskip}%
  }%
  {\protect\endProof%
   \protect\fussy%
  }
\makeatother

\makeatletter
\gdef\tabbing{\lineskip \z@skip\let\>\@rtab\let\/\@ltab
  \let\=\@settab
     \let\+\@tabplus\let\-\@tabminus\let\`\@tabrj\let\'\@tablab
     \let\\=\@tabcr
     \@hightab\@firsttab
     \global\@nxttabmar\@firsttab
     \dimen\@firsttab\@totalleftmargin
     \global\@tabpush\z@ \global\@rjfieldfalse
     \trivlist \item\relax
     \if@minipage\else\vskip\parskip\fi
     \setbox\@tabfbox\hbox{%
       \rlap{\hskip\@totalleftmargin\indent\the\everypar}}%
     \def\@itemfudge{\box\@tabfbox}%
     \@startline\ignorespaces}
\makeatother

\floatstyle{ruled}
\newfloat{tabl}{tbp}{lot}
\floatname{tabl}{Table}
\newrefformat{tab}{Table~\ref{#1}}
\newlength{\tablength}
  {\begin{tabl}[#1]%
   \caption{%
     \hspace*{\fill}#3\hspace*{\fill}
     \settowidth{\tablength}{\textbf{Table \thetable}}\hspace*{\tablength}}%
   \label{#2}%
   \normalsize}%
  {\end{tabl}}
\newenvironment{tab*}[3][]%
  {\begin{tabl*}[#1]%
   \caption{%
     \hspace*{\fill}#3\hspace*{\fill}
     \settowidth{\tablength}{\textbf{Table \thetable}}\hspace*{\tablength}}%
   \label{#2}%
   \normalsize}%
  {\end{tabl*}}
\newfloat{system}{ptbh}{los}
\floatname{system}{System}
\newrefformat{sys}{System~\ref{#1}}
  {\begin{system}%
   \caption{%
     \hspace*{\fill}#2\hspace*{\fill}
     \settowidth{\tablength}{\textbf{Table \thetable}}\hspace*{\tablength}}%
   \label{#1}%
   \normalsize}%
  {\end{system}}

\def\timestamp{}
\def\Time-stamp: <#1>{\gdef\timestamp{#1}}

\newcommand{\li}{\lstinline}

\newsavebox{\ceqbox}\savebox{\ceqbox}{$:=\ $}
\newsavebox{\neqbox}\savebox{\neqbox}{$\neq$}
\newsavebox{\eqbox}\savebox{\eqbox}{$=$}

\lstdefinelanguage{llw}{
  columns=flexible,
  basicstyle=\tt,
  identifierstyle=\it,
  emphstyle=\rm,
  emph={nil,NULL},
  alsoletter={'},
  morekeywords={new, dispose, else, if, local, while},
  mathescape=true,
  showlines=true,
  literate={{->}{$\shortrightarrow$}1%
            {!=}{{\usebox{\neqbox}}}2%
            {=}{{\usebox{\ceqbox}\ }}2%
            {==}{{\usebox{\eqbox}}}2}
}
\makeatletter\@ifundefined{theorem}{

}\makeatother

\newrefformat{thm}{Theorem~\ref{#1}}
\newrefformat{prop}{Proposition~\ref{#1}}
\newrefformat{lem}{Lemma~\ref{#1}}
\newrefformat{cor}{Corollary~\ref{#1}}
\newrefformat{conj}{Conjecture~\ref{#1}}
\newrefformat{obs}{Observation~\ref{#1}}
\newrefformat{defn}{Definition~\ref{#1}}
\newrefformat{algo}{Algorithm~\ref{#1}}

\newrefformat{appendix}{Appendix~\ref{#1}}

\newif\iflong
\longtrue

\makeatother  

\semanticdomain{Val[Values], Loc, Var[Variables], Fields}
\proglangkeyword{xor[\,xor\,], local[local\;], new, disp[dispose], nondet}
\syntacticpredicate{jsr, tree,list,lseg,dlseg,xlseg,true,false}
\semanticvalue{fault, nil}
\metalangfunc{post, owned, vars[var], req, mod, chop, vcg, er, par, cstr, ptsto, nf, unroll}
\acronym{ccr,vc,dag,deq}
\newcommand{\pvs}{\ps;;\vs}

\newcommand{\Es}{\vec{E}}

\newcommand{\ps}{\vec{p}}
\newcommand{\rs}{\vec{r}}
\newcommand{\us}{\vec{u}}
\newcommand{\vs}{\vec{v}}
\newcommand{\xs}{\vec{x}}
\newcommand{\ys}{\vec{y}}
\newcommand{\zs}{\vec{z}}
\newcommand{\Ss}{\mathit{SI}}
\newcommand{\xfm}[3][]{\spec{#2}{\<jsr>\ifthenelse{\equal{#1}{}}{}{_{#1}\,}}{#3}}

\renewcommand{\disjoint}[2]{#1\cap#2=\emptyset}
\renewcommand{\spec}[3]{[#1]\,#2\,[#3]} 
\mathlig{;}{\loPrecLogConSpc{\wedge}} 

\infrule{Assign, Conditional, Dispose, Empty, Lookup, Mutate, New,
  SwitchE[Switch$(E)$], UnrollListsegE[Unroll List Segment$(E)$],
  UnrollTreeE[Unroll Tree$(E)$], StarI[$*$-Introduction],
  incy[Inconsistency], subst[Substitution], eql[$==$-L],
  nnlv[NilNotLVal], starp[$*$-Partial], exm[ExcludedMiddle], EmpTreeL,
  EmpLsL, ax[Axiom], eqr[$==$-R], hyp[Hypothesis], ident[Identity],
  exist[Existential], EmpTreeR, EmpLsR, RollTree, RollLs, AppendLsNil,
  AppendLsGuard, IndPredL, IndPredR}

\infrule{eq[Existential Quantification], const[Constancy],
  conj[Conjunction], conseq[Consequence], frame[Frame], cf[Constancy
  \& Frame], seq[Sequential Composition], pc[Procedure Call],
  vs[Variable Substitution], ps[Parameter Substitution]}

\renewcommand{\proglangkeywordstyle}[1]{\smtt{#1}}
\newcommand{\identstyle}[1]{{\normalsize\it#1}}
\lstdefinelanguage{sf}{
  columns=flexible,
  emphstyle=\small\tt,
  basicstyle=\small\tt,
  identifierstyle=\identstyle,
  emph={nil,NULL},
  alsoletter={'},
  morekeywords={new, dispose, else, if, local, while, with, when, resource, xor, skip, nondet},
  mathescape=true,
  showlines=true,
  literate={{->}{$\shortrightarrow$}1%
            {=}{{{\usebox{\ceqbox}}}}2%
            {||}{{$\parallel\ $}}2%
            {NULL}{\semanticvaluestyle{nil}}3%
            {!=}{{\usebox{\neqbox}}}2%
            {==}{{\usebox{\eqbox}}}2}
}
\renewcommand{\li}[1]{\lstinline!#1!}

\renewcommand{\ite}[3]{\plks{if}(#1)\;\{#2\}\;\plks{else}\;\{#3\}}
\renewcommand{\while}[3][]{
 \plks{while}(#2)\ifthenelse{\equal{#1}{}}{}{\;[#1]\;}\{#3\}}
\renewcommand{\with}[3]{\plks{with}\;#1\;\plks{when}(#2)\;\{#3\}}

\title{Verification Condition Generation and\\Variable Conditions in Smallfoot
}

\author{Josh~Berdine\inst{1}
   \and Cristiano~Calcagno\inst{2}
   \and Peter~W.~O'Hearn\inst{3}}
\institute{Microsoft Research
      \and ETH Zurich, Imperial College London, and Monoidics Ltd
      \and University College London}

\begin{document}
\maketitle

\begin{abstract}
  These notes are a companion to \cite{BerdineCO05c} which describe
  \begin{itemize}
  \item the variable conditions that Smallfoot checks,
  \item the analysis used to check them,
  \item the algorithm used to compute a set of verification conditions
    corresponding to an annotated program, and
  \item the treatment of concurrent resource initialization code.
  \end{itemize}
\end{abstract}

\section*{2012 Introduction}

This document presents the variable conditions and checking algorithms as
implemented in Smallfoot 0.1 of late 2005.  These conditions on the proof
rules for concurrency rely on some of the relaxations introduced by Brookes
\cite{Brookes07} relative to O'Hearn's system \cite{OHearn07}, and so
originally soundness was via Brookes's result.  Ian Wehrman and Berdine have
since found some cases where these relaxed conditions are unsound, prompting a
revisitation of this topic.  Brookes \cite{Brookes11} and Reddy and Reynolds
\cite{ReddyR12} have recently introduced systems which address these issues
while admitting encodings of proofs in O'Hearn's more restrictive system
(among other improvements).

In hindsight, while Smallfoot needs more than O'Hearn's system, it does not
use the full relaxation of Brookes's original system, in particular retaining
concurrency condition 5 below.  As a result, the proofs found by Smallfoot
appear to be embeddable into either of the recently proposed sound systems,
although we do not claim to have a formal proof at this time.

The condition checking algorithms remain non-compositional, and hence uses a
whole-program analysis, despite the compositionality of both Brookes's revised
system and that of Reddy and Reynolds.  This is a result of the necessity of
guiding the search for a proof to one which satisfies the occurrence
conditions, as opposed to the distinct problem of checking whether a given
candidate proof outline satisfies the conditions, or of inferring a valid
permissions annotation as in \cite{ReddyR12}.

\section{Checking Variable Conditions}

\subsection{Annotated Programs}

Each Smallfoot program determines a resource environment @G which contains the
resource declarations
$$r_i(\xs_i)R_i$$
where $\xs_i$ and $R_i$ are resource $r_i$'s protected variables and
invariant; and a procedure environment @D which contains the procedure
declarations
$$f(\pvs) \spec{P_f}{C_f}{Q_f}$$
where procedure $f$'s parameters $\ps$ are passed by reference and $\vs$ by
value, and assertions $P_f$ and $Q_f$ are $f$'s pre- and post-conditions.  We
assume that @G and @D are given.

Commands are generated by:
\begin{eqnarray*}[rclqTlql]
  E &::=& x | \<nil> | c | E \<xor> E \\
  B &::=& E==E | E!=E \\
  S &::=& x:=E | x:=E->t | E->t:=E | x:=\<new>() | \<disp>(E) \\
  C &::=& S | C;.C | \ite{B}{C}{C} | \while[I]{B}{C} \\
      &|& f(\xs;;\Es) | f(\xs;;\Es)||f(\xs;;\Es) | \with{r}{B}{C}
\end{eqnarray*}

\subsection{Legal Annotated Programs}

Using the following notation
\begin{eqnarray*}
  \<owned>(\rs) &\Def=& \bigcup_{i.r_i\in\rs}\xs_i \\
  \<vars>(\rs) &\Def=& \<owned>(\rs) \u \bigcup_{i.r_i\in\rs}\<fv>(R_i)
\end{eqnarray*}
where $\rs$ denotes a set of resources, the set of legal annotated
programs is restricted by the following constraints:
\begin{itemize}
\item In any procedure call $f(\ys;;\Es)$ or region $\with{r}{B}{C}$ the
  variable $f$/$r$ must be defined in a procedure/resource declaration.
\item In every procedure declaration $f(\pvs) \spec{P_f}{C_f}{Q_f}$,
  the formal parameters $\ps,\vs$ are all distinct.
\item The resources $\rs$ must be distinct.
\item The protection lists must all be disjoint:
  $\disjoint{\<owned>(r_i)}{\<owned>(r_j)}$ when $i \neq j$.
\item No resource's invariant may have a free occurrence of a variable
  in a distinct resource's protection list:
  $\disjoint{\<fv>(R_i)}{\xs_j}$ when $i \neq j$.
\end{itemize}

\subsection{Simplifying Assumptions}

Additionally, we assume a pre-processing phase which renames bound
variables to satisfy the following simplifying assumptions:
\begin{itemize}
\item Bound variables (formal parameters and variables bound by
  \<local>@!) are distinct from one another, and from global
  variables.
\item For each procedure declaration, variables in the postcondition
  are not bound (by \<local>@!) in the body.
\end{itemize}

\subsection{Variable Conditions}

We define functions $\<vars>(--),\<mod>(--),\<req>(--)$ on commands
$C$ and procedure names $f$.  Intuitively:
\begin{itemize}
\item $\<vars>(C)$ is set of variables that $C$ mentions (in the
  program or specifications, recursively) without protection;
\item $\<mod>(C)$ is the set of variables that $C$ may modify without
  protection (acquiring $r$ protects $\<owned>(\set{r})$);
\item $\<req>(C)$ is the set of resources required to be acquired
  before executing $C$.
\end{itemize}
The auxiliary function $\<er>(M,A)$, returning the set of resources
that need to be acquired before modifying variables in $M$ and
accessing variables in $A$, is defined as:
\begin{eqnarray*}
  \<er>(M,A) &=& \Bigl\{r \ \Bigr\vert\
  \begin{eqnalign}[l]
    r(\xs)R\in@G \text{ and } \\
    (A \n \xs \not= \emptyset \text{ or } M \n (\xs \u \<fv>(R)) \not= \emptyset)
  \end{eqnalign}\Bigr\}
\end{eqnarray*}
The definition proceeds by performing a simple fixpoint calculation to
determine the least solution (which is best) of the following
equations, ordering by point-wise subset inclusion of functions:
\begin{eqnarray*}[rcl]
\<vars>(f) &=& (\<vars>(C) \u \<fv>(P,Q))-(\ps \u \vs) \\
\<mod>(f) &=& \<mod>(C)-(\ps \u \vs) \\
\<req>(f) &=& \<req>(C) \u \<er>(\emptyset,\<fv>(P,Q)-(\ps \u \vs))
\end{eqnarray*}
where $f(\pvs)\spec{P}{C}{Q} \in @D$, and for commands:
\begin{eqnarray*}[rcl]
\<vars>(x:=E) &=& \set{x} \u \<fv>(E) \\
\<vars>(x:=E->t) &=& \set{x} \u \<fv>(E) \\
\<vars>(E->t:=F) &=& \<fv>(E) \u \<fv>(F) \\
\<vars>(x:=\<new>()) &=& \set{x} \\
\<vars>(\<disp>(E)) &=& \<fv>(E) \\
\<vars>(C;.C') &=& \<vars>(C) \u \<vars>(C') \\
\<vars>(\ite{B}{C}{C'}) &=& \<fv>(B) \u \<vars>(C) \u \<vars>(C') \\
\<vars>(\while[I]{B}{C}) &=& \<fv>(I,B) \u \<vars>(C) \\
\<vars>(f(\xs;;\Es)) &=& \<vars>(f) \u \xs \u \<fv>(\Es) \\
\<vars>(f(\xs;;\Es)||f'(\xs';;\Es')) &=& \<vars>(f(\xs;;\Es)) \u \<vars>(f'(\xs';;\Es')) \\
\<vars>(\with{r}{B}{C}) &=&
((\<fv>(B) \u \<vars>(C)) - \<fv>(R_r)) \u (\<mod>(C) - \<owned>(\{r\})) \\
\\
\<mod>(x:=E) &=& \set{x} \\
\<mod>(x:=E->t) &=& \set{x} \\
\<mod>(E->t:=F) &=& \emptyset \\
\<mod>(x:=\<new>()) &=& \set{x} \\
\<mod>(\<disp>(E)) &=& \emptyset \\
\<mod>(C;.C') &=& \<mod>(C) \u \<mod>(C') \\
\<mod>(\ite{B}{C}{C'}) &=& \<mod>(C) \u \<mod>(C') \\
\<mod>(\while[I]{B}{C}) &=& \<mod>(C) \\
\<mod>(f(\xs;;\Es)) &=& \<mod>(f) \u \xs \\
\<mod>(f(\xs;;\Es)||f'(\xs';;\Es')) &=& \<mod>(f(\xs;;\Es)) \u \<mod>(f'(\xs';;\Es')) \\
\<mod>(\with{r}{B}{C}) &=& \<mod>(C) - \<owned>(\{r\}) \\
\\
\<req>(S) &=& \<er>(\<mod>(S),\<vars>(S)) \\
\<req>(C;.C') &=& \<req>(C) \u \<req>(C') \\
\<req>(\ite{B}{C}{C'}) &=& \<req>(C) \u \<req>(C') \u \<er>(\emptyset,\<fv>(B)) \\
\<req>(\while[I]{B}{C}) &=& \<req>(C) \u \<er>(\emptyset,\<fv>(I,B)) \\
\<req>(f(\xs;;\Es)) &=& \<req>(f) \u \<er>(\xs,\<fv>(\Es)) \\
\<req>(f(\xs;;\Es)||f'(\xs';;\Es')) &=& \<req>(f(\xs;;\Es)) \u \<req>(f'(\xs';;\Es')) \\
\<req>(\with{r}{B}{C}) &=& (\<req>(C) \u \<er>(\emptyset,\<fv>(B))) -\{r\}
\end{eqnarray*}

The definitions of \<vars> and \<mod> are as expected, except that \<ccr>
statements $\with{r}{B}{C}$ hide accesses and modifications of the variables
$\<owned>(r)$ in $C$.  Note that variables in $\<fv>(R_r)$ can only be
modified in a critical region for $r$. Therefore, when computing the external
effect of a command $\with{r}{B}{C}$ we can ignore the reads to $\<fv>(R_r)$
since they can never happen in parallel with a write.  For \<req>, any mention
of a variable protected by $r$, or modification of a variable in $r$'s
invariant, causes $r$ to be required; and a \<ccr> for $r$ discharges the
requirement of $r$.

\subsubsection{Variable Aliasing Conditions}

The variable conditions for aliasing follow those of \cite{Hoare71} and
\cite{Cook78}.  The conditions needed to avoid variable (not heap) aliasing
are enforced by checking, for every procedure call $f(\xs;;\Es)$ in the
program:
\begin{itemize}
\item The actual reference parameters $\xs$ are distinct.
\item If a global variable $z$ is passed by reference, then $f$ and procedures
  $f$ calls, recursively, must not read or modify $z$, or mention it in
  specifications: $\disjoint{\xs}{\<vars>(f)}$.
\end{itemize}

\subsubsection{Variable Conditions for Concurrency}

The variable conditions for concurrency follow those of
\cite{OwickiGries76,OHearn07,Brookes07}.  The first two concurrency
conditions:
\begin{enumerate}
\item[1.] Protected variables of $r$ appear only within \<ccr>s for
  $r$.
\item[2.] Variables appearing in a resource $r$'s invariant can only
  be modified within \<ccr>s for $r$.
\end{enumerate}
are checked using the computed $\<req>(--)$.  A violation of one of
these conditions results in ``too large'' a required resources set for
the offending code, which eventually propogates to the main procedure.
Hence, we check that $\<req>(\smtt{main}) = \emptyset$, if it appears.
Note that this analysis ignores deadlock due to acquiring an
already-held resource, as Smallfoot only proves safety.

The third and fourth concurrency conditions:
\begin{enumerate}
\item[3.] Only protected variables can be modified in one parallel
  process and read or mentioned in specifications in another.
\item[4.] For each parallel composition $f(\xs;;\Es)||f'(\xs';;\Es')$,
  $f(\xs;;\Es)$ and the specification of $f$ cannot mention variables
  modified by $f'(\xs';;\Es')$, and vice versa.
\end{enumerate}
are checked using the computed $\<vars>(--)$ and $\<mod>(--)$:
\begin{eqnarray*}[Trql]
  & \disjoint{\<mod>(f(\xs;;\Es))}{(\<fv>(P',Q') \u \<vars>(f'(\xs';;\Es')))}
  \\and
  & \disjoint{\<mod>(f'(\xs';;\Es'))}{(\<fv>(P,Q) \u \<vars>(f(\xs;;\Es)))}
  ,\\where
  & f(\pvs)\spec{P}{C}{Q},f'(\ps';;\vs')\spec{P'}{C'}{Q'} \in @D
  .
\end{eqnarray*}

The final concurrency condition is a property of program proofs, not of
annotated programs themselves:
\begin{enumerate}
\item[5.] Whenever a \<ccr> is symbolically executed, the pre and post
  states cannot mention variables modified by other processes.
\end{enumerate}
Note that according to the inference rule for \<ccr>s, entering a
\<ccr> adds the resource invariant to the current precondition.  Also,
it may be that processes running in parallel with the one executing
the \<ccr> under consideration modify variables appearing in the added
invariant.  For this reason, we introduce a further analysis which
computes, for each procedure $f$, the set $\<par>(f)$ of procedures
that might run in parallel with $f$:
\begin{enumerate}
\item $\<par>(f) \supseteq \{f'\}$ for all occurrences of
  $f(\xs;;\Es)||f'(\xs';;\Es')$ or $f'(\xs';;\Es')||f(\xs;;\Es)$ in
  the program;
\item $\<par>(f') \supseteq \<par>(f)$ for all occurrences of
  $f'(\xs';;\Es')$ or $f'(\xs';;\Es')||f''(\xs'';;\Es'')$ or
  $f''(\xs'';;\Es'')||f'(\xs';;\Es')$ in $C_f$.
\end{enumerate}
(As before, we take the smallest set satisfying the above conditions.)
The results of this analysis are then used in VCGen to instrument the
\<vc>s for \<ccr>s so that any such variables modified by processes in
parallel are quantified out of the post states of \<ccr>s during
symbolic execution, thereby avoiding bad proofs.

Checking the concurrency conditions essentially classifies each variable into
one of the following five classes:
\begin{description}
\item[Local] variables are declared by \<local>@!, or are procedure value
  parameters.  Their use is unrestricted within their scope.
\item[Process-local] variables appear, and are mentioned in specifications, in
  only one process, and do not appear in any resource invariants.  In that
  process, their use is unrestricted.
\item[Global-constant] variables appear in some function and are not local.
  They cannot be written to but can be read or appear in specifications,
  including resource invariants, in any process.
\item[Protected] variables are those which appear in one resource's protection
  list, and can be modified, accessed, or mentioned in specifications in any
  process, but only within critical regions for the associated resource.
\item[Process-protected] variables appear in at least one resource invariant
  and in only one process.  In that process they are modified only within
  critical regions for all the resources in whose invariants they appear.
  Also in that process, they can be read and appear in specifications outside
  of critical regions.  Variables which appear free in some resource invariant
  but are not protected are either process-protected or global-constant,
  depending on whether they are ever written to.
\end{description}

\section{Verification Condition Generation}

\subsection{Verification Conditions}

A verification condition is a triple
$
\spec{P}{\Ss}{Q}
$
where $\Ss$ is a ``symbolic instruction'':
\begin{eqnarray*}[rclqTlql]
  \Ss &::=& @q | S | \xfm[\xs]{P}{Q} | \itee{B}{\Ss}{\Ss} | \Ss;.\Ss
\end{eqnarray*}
A symbolic instruction is a piece of loop-free sequential code where
all procedure calls have been instantiated to $\<jsr>$ instructions of
the form $\xfm[\xs]{P}{Q}$.  This form plays a central role in
Smallfoot. We use it not only to handle procedure calls, but also for
concurrency and for entry to and exit from a critical region.

Semantically, $\xfm[\xs]{P}{Q}$
is a ``generic command'' in the sense of \cite{Schwarz77}.
It is the greatest relation satisfying the pre- and post-condition, and subject to the constraint that
only the variables in $\xs$ are modified.

The symbolic execution rule for the \<jsr> instruction is:
\[
  \drule[\xs' \<fresh>]{}
        {@P/|@S |- @P' /| @S' * @S_F\\
         \spec{(@P[\xs'/\xs]) /| Q * (@S_F[\xs'/\xs])} {C} {R}}
        {\spec{@P /| @S} {(\xfm[\xs]{@P'/|@S'}{Q} ;. C)} {R}}
\]
To apply this rule we have to discover a frame axiom $@S_F$ which
describes the portion of heap unchanged by a procedure call, and
\cite{BerdineCO05} describes a proof-theoretic method for obtaining
them.

\subsection{VCGen}
\label{vcgensec}

For each procedure declaration $f(\pvs)\spec{P}{C}{Q}$ we generate a set of
verification conditions $\<vcg>(f, \spec{P}{C}{Q})$, which is itself defined
using a helper function \<chop> that takes a command and produces a symbolic
instruction together with a set of verification conditions.  \<vcg> just runs
\<chop> on the body $C$, tacks the pre and post onto the resulting symbolic
instruction, and adds that to the verification conditions returned by \<chop>.

  \begin{eqnarray*}
    \<vcg>(g, \spec{P}{C}{Q}) &=&
    \set{\spec{P}{\Ss}{Q}} \u L \\&&
      \text{ where } \Ss ,, L = \<chop>(g, C)
  \\[1ex]
    \<chop>(g, S) &=& S,, \emptyset
    \\
    \<chop>(g, C;.C') &=&
      \Ss ;. \Ss' ,, L' \u L' \\&&
      \text{where }
        \Ss,, L = \<chop>(g, C) \text{ and }
        \Ss' ,, L' = \<chop>(g, C')
    \\
    \<chop>(g, \ite{B}{C}{C'}) &=&
      \itee{B}{\Ss}{\Ss'},, L \u L' \\&&
      \text{where }
        \Ss ,, L = \<chop>(g, C) \text{ and }
        \Ss' ,, L' = \<chop>(g, C')
    \\
    \<chop>(g, \while[I]{B}{C}) &=&
      \xfm[\<mod>(C)]{I}{!B /|I},, \<vcg>(\spec{B /| I}{C}{I})
    \\
    \<chop>(g, f(\xs;;\Es)) &=&
      (\xfm[\emptyset]{\<emp>}{\vs'==\Es;\<emp>} ;.
       \xfm[{\<mod>(C_f) [\xs/\ps,\vs'/\vs]}]{P[\xs/\ps,\vs'/\vs]}{Q[\xs/\ps,\vs'/\vs]})
      ,, \emptyset \\&&
      \text{where }
        f(\pvs) \spec{P}{C}{Q} \text{ and } \vs' \text{ fresh}
    \\
    \<chop>(g, f(\xs;;\Es)||f'(\xs';;\Es')) &=&
      (\xfm[\emptyset]{\<emp>}{\vs==\Es /| \vs'==\Es';\<emp>} ;.
       \xfm[\zs\u\zs']{P*P'}{Q*Q'}),,\emptyset \\&&
      \begin{eqnalign} \text{where }
        \xfm[\emptyset]{\<emp>}{\vs==\Es;\<emp>} ;. \xfm[\zs]{P}{Q},, \emptyset
          &=& \<chop>(g, f(\xs;;\Es)) \\
        \xfm[\emptyset]{\<emp>}{\vs'==\Es';\<emp>} ;. \xfm[\zs']{P'}{Q'},, \emptyset
          &=& \<chop>(g, f'(\xs';;\Es'))
      \end{eqnalign}
    \\
    \<chop>(g, \with{r}{B}{C}) &=&
      \bigl(\xfm[\emptyset]{\<true>;\<emp>}{B;R} ;. \Ss ;.
            \xfm[\xs\u\us]{\<true>;R}{\<true>;\<emp>}\bigr) ,, L \\&&
      \text{where }
        \Ss,, L = \<chop>(g, C) \text{ and } r(\xs)R \text{ and }
        \us = \<fv>(R) \n \textstyle\bigcup_{f \in \<par>(g)} \<mod>(f)
  \end{eqnarray*}

The definition of \<chop> for primitive statements, sequential
composition, conditionals and loops is mostly as expected, except that
for loops we generate a \<jsr> instruction that allows invariants to
be smaller than they might otherwise be, because of framing.

For procedure call, we rename the value parameters and use two
\<jsr>'s: the first to initialize the renamed parameters and the
second to abstract the body of the procedure, using only its spec.
This renaming allows the postcondition to refer to the initial value
of the parameters which are not modified by the body. The composition
of the two \<jsr>'s satisfies a spec $\spec{A}{-}{B}$ iff the second
one satisfies $\spec{A /| \vs'==\Es}{-}{B}$.  In the definition,
$\<mod>(C_f)$ is the set of variables modified by $C_f$ (or one of the
procedures that $C_f$ calls) except for protected variables modified
within a \<ccr>.

For parallel composition we emit two \<jsr>'s that combine the
initializations of the two procedure calls, and take the
$*$-combinations of the respective preconditions and postconditions,
following the parallel proof rule
$$
\drule[]{}{\spec{P}{C}{Q} \qquad \spec{P'}{C'}{Q'}}
{\spec{P*P'}{C\parallel C'}{Q*Q'}}
.$$

Entry to, and exit from, \<ccr>s is modeled by \<jsr> instructions.
The entry \<jsr> adds the resource invariant and boolean condition to
the symbolic state.  Since this represents adding \emph{any} concrete
heap satisfying the invariant and condition, upon entry to a \<ccr>
the body cannot assume or depend on anything further about the
acquired heap.  This is how we handle potential interference from
parallel processes, which may change one concrete heap satisfying the
invariant to another.  Additionally, outside interference is prevented
in code following a \<ccr> since the exit \<jsr> removes the resource
invariant from the symbolic state and forgets the values of variables
which are protected, $\xs$, or might be modified by processes
$\<par>(f)$ running in parallel, $\us$.  The net result is that
correctness of a parallel program is reduced to several sequential
triples, and no interleaving needs to be considered.  This \<vc>
definition follows the description of the \<ccr> proof rule
$$
\drule[]{}
{\spec{(P*R_r)\wedge B}{C}{Q*R_r}
}
{\spec{P}{\with{r}{B}{C}}{Q}
}
$$
(where $R_r$ is an invariant formula associated with resource $r$) and both
occurrences of \<jsr> make use of the frame axiom inference capability; the
precondition $P$ of a \<ccr> is maintained after the entry \<jsr>, and an
appropriate $Q$ part for the postcondition in the rule is discovered as a
frame axiom for the exit \<jsr>.  The $r(\xs)R$ in the where clause indicates
that $R$ is the declared invariant of $r$ in the program.

To tie all of this together there is one further check that must be
made.  The \li{init} procedure must establish all of the resource
invariants, separately, and the precondition of \li{main} if a main
procedure is included.  So given $\mathtt{init}()\spec{P}{C}{Q}$ and
$\mathtt{main}()\spec{P'}{C'}{Q'}$ we check the entailment $Q|- R_1*
\cdots * R_n*P'$.  We also require $C$ to not contain procedure calls,
\<ccr>s, or parallel compositions.  All told, the property that this
establishes (following the rule for complete programs
\cite{OHearn07,Brookes07}) for a program is
$$
\spec{P} {C;. \mbox{RESDECLS};. \<let> \mbox{PROCDECLS} \<in> C'} {Q'*R_1*\cdots *R_n}
$$
where PROCDECLS consists of those procedure declarations other than  \li{main}
and  \li{init}.
(We could also include a finalization procedure that disposes of the $R_i$ at the end.)

\section{Resource Initialization}

Resource initializers are subject to the following constraints:
\begin{enumerate}
\item No resource's initializer modifies a variable mentioned by a
  distinct resource: $\disjoint{\<mod>(C_i)}{\<vars>(r_j)}$ when $i
  \neq j$.

  This is performed by checking, for all $i$
  \begin{eqnarray*}[Trql]
  & \disjoint{\<mod>(C_i)}{\<vars>(C_1,.,C_{i-1})} \\
  and & \disjoint{\<vars>(C_i)}{\<mod>(C_1,.,C_{i-1})}
  \end{eqnarray*}
\item The $\smtt{init}$ procedure, if it appears, and the resource
  initializers $C_i$ contain no procedure calls or \<ccr>s.
\end{enumerate}

These constraints ensure that the order in which the initializers are
executed is immaterial.  Therefore, we have an additional verification
condition:
\[
\<vcg>(\spec{P}{C}{Q * R_1 * \cdots * R_n})
\]
where $R_1, \ldots, R_n$ are all the resource invariants and
\[
P,C,Q =
\begin{cases}
  P',C',Q' & if $\smtt{init}()\spec{P'}{C'}{Q'} \in @D$ \\
  \<emp>, C_1 ;. \cdots ;. C_n, \<emp> \ & otherwise
\end{cases}
\]
where $C_1, \ldots, C_n$ are all the resource initializers (in some
unspecified order).

Also, in case $\smtt{init}$ appears, the precondition of procedure
$\smtt{main}$, if it appears, is taken to be the postcondition of
$\smtt{init}$, irrespective of what appears in the file.

\bibliographystyle{abbrv}
\bibliography{db-short}

\end{document}